\documentclass[12pt]{article}

\usepackage{graphics}
\usepackage{epsfig}
\usepackage{cite}
\input epsf

\usepackage{amssymb,amsmath,euscript}
\usepackage{feynmf}
\usepackage{indentfirst}
\usepackage{cite}

\title{Drell-Yan lepton pair production at high energies \\ in the $k_T$-factorization approach}

\author{A.V.\,Lipatov, M.A.\,Malyshev, N.P.\,Zotov}

\begin{document}

\maketitle


\begin{center}
{\it D.V.~Skobeltsyn Institute of Nuclear Physics,\\ 
M.V. Lomonosov Moscow State University,
\\119991 Moscow, Russia\/}\\[3mm]
\end{center}

\vspace{0.5cm}

\begin{center}

{\bf Abstract }

\end{center}

In the framework of the $k_T$-factorization approach, the production of unpolarized 
Drell-Yan lepton pair at high energies is studied. The consideration is based
on the ${\cal O}(\alpha)$ and ${\cal O}(\alpha \alpha_s)$ off-shell partonic matrix elements
with virtual photon $\gamma^*$ and $Z$ boson exchange. 
The calculations include leptonic decays of $Z$ bosons 
with full spin correlations as well as $\gamma^* - Z$ interference. 
The unintegrated parton densities in a proton are determined by the 
Kimber-Martin-Ryskin prescription. 
Our numerical predictions are compared with the data taken by the D$\emptyset$, CDF and CMS
collaborations at the Tevatron and LHC energies.
Special attention is put on the specific 
angular distributions measured very recently by the CDF collaboration
for the first time.

\vspace{0.8cm}

\noindent
PACS number(s): 12.38.-t, 12.15.Ji

\vspace{0.5cm}

\section{Introduction}

With the start of the LHC experiments, high energy particle physics entered a new era.
The LHC opens a new kinematic regime where the number of novel physical
phenomena can occur. One of the most useful tools to study
hadronic interactions at high energies is the Drell-Yan dilepton production, in
which quark-antiquark annihilation form intermediate virtual photon or $Z$ bosons decaying to lepton pairs.
This process is presently of considerable interest from both experimental and theoretical points of view.
In particular, Drell-Yan pair production is a unique process which offers high
sensitivity to the parton (quark and gluon) distributions in a proton.
It provides a major source of background
to a number of processes such as Higgs, $t \bar t$ pair, di-boson or $W'$ and $Z'$ bosons production 
(and other processes beyond the SM) studied at hadron colliders.
Dilepton production has 
a large cross section and clean signature
in the detectors and therefore it is used for monitoring of the collider luminosity and calibration of
detectors. Moreover, it is an important reference process for measurements of electroweak boson properties at hadron
colliders. Therefore it is essential to have an accurate QCD predictions for 
corresponding cross sections and related kinematical distributions.

Theoretical investigations of Drell-Yan pair production have own long story.
It is one of the few processes in hadron-hadron collisions where
the collinear QCD factorization has been rigorously proven\cite{1,2,3,4}. Within this framework, 
the NLO pQCD calculations of inclusive cross sections have been performed\cite{5,6,7}, 
and later it was done on up to NNLO accuracy\cite{8,9}.
Recently fully exclusive NNLO pQCD calculations became available, including the leptonic decay of intermediate $Z$ boson\cite{10,11,12,13}. 
The results of these calculations
agree with the Tevatron and LHC data within the theoretical and experimental uncertainties.
Of course, typically for collinear QCD factorization, perturbative calculations diverge at small dilepton 
transverse momenta $p_T \ll M$ (where $M$ is the invariant mass of produced lepton pair) 
with terms proportional to $\ln M/p_T$ appearing due to soft and collinear gluon emission.
Therefore special soft gluon resummation technique\cite{14,15,16,17,18,19,20} should be used to make QCD predictions at low $p_T$. 
Such soft gluon resummation can be performed either in the
transverse momentum space\cite{21} or in the Fourier conjugate impact parameter space\cite{22}.
Differences between the two formalisms are discussed in\cite{23}.
The traditional calculations combine fixed-order perturbation theory with analytic resummation
and some matching criterion. 

An alternative description can be achieved within the framework of the $k_T$-factorization approach of QCD\cite{24}.
This approach is based on the Balitsky-Fadin-Kuraev-Lipatov (BFKL)\cite{25} or 
Ciafaloni-Catani-Fiorani-Marchesini (CCFM)\cite{26} equations and 
provides solid theoretical grounds for the effects of initial gluon radiation and intrinsic
parton transverse momentum $k_T$.
A detailed description and discussion of the $k_T$-factorization formalism can be
found, for example, in reviews\cite{27}. Here we only mention that,
in contrast with the collinear approximation of QCD, the initial gluon emissions 
in the $k_T$-factorization approach generate the finite dilepton transverse momentum $p_T$
already at Born level.
Moreover, the soft gluon resummation formulas are the result of the 
approximate treatment of the solutions of CCFM equation, as it was shown in\cite{28}.

In the present note we apply the $k_T$-factorization approach to unpolarized Drell-Yan pair
production in $p\bar p$ and $pp$ collisions.
A non-collinear factorization theorem for this process has been proven\cite{29} for $p_T \ll M$.
Below we assume it in a wide range of $p_T$ for phenomenological 
purposes\footnote{For discussion  of the $k_T$-factorization for 
high energy processes see, for example,\cite{30}.}.
First application of $k_T$-factorization approach to lepton pair production has been performed in\cite{31},
where authors have considered only diagrams with virtual photon exchange and
concentrated mostly on the rather low energies covered by the RHIC and UA1 experiments.
More general consideration of high energy resummation for Drell-Yan processes was done in\cite{32}.
Our main goal is to give a systematic analysis of the
Tevatron data\cite{33,34,35,36,37,38} and first LHC measurements\cite{39} performed by the CMS collaboration.
The consideration is based on the ${\cal O}(\alpha)$ and ${\cal O}(\alpha \alpha_s)$ 
off-shell (depending on the non-zero transverse momenta of incoming partons) production 
amplitudes where we take into account both $\gamma^*$ and $Z$ boson exchange.
Our calculations include also leptonic decays of $Z$ bosons 
with full spin correlations. Thus we easily can produce various kinematical distribution and 
apply cuts, analogous to experimental ones.
Specially we study the angular distributions of produced lepton pair
measured very recently\cite{37} by the CDF collaboration for the first time
and investigate the different sources of theoretical uncertainties.

The outline of our paper is following. In Section~2 we 
recall shortly the basic formulas of the $k_T$-factorization approach with a brief 
review of calculation steps. In Section~3 we present the numerical results
of our calculations and a discussion. Section~4 contains our conclusions.

\section{Theoretical framework} \indent

There are three subprocesses which describe Drell-Yan pair production at order of 
${\cal O}(\alpha)$ and ${\cal O}(\alpha \alpha_s)$: 
\begin{gather}
\label{qq2ll}
q + \bar q \to \gamma^*/Z \to l^+ + l^-\\
\label{qg2llq}
q + g^* \to \gamma^*/Z + q\to l^+ + l^- + q\\
\label{qq2llg}
q + \bar q \to \gamma^*/Z + g \to l^+ + l^- + g
\end{gather}

\noindent
The corresponding Feynman graphs are shown on the Fig.~1.
Note that in the framework of $k_T$-factorization approach 
contribution from the subprocess~(3) 
is already taken into account by the quark-antiquark annihilation~(1) due to the 
initial state gluon radiation.
Therefore below to avoid the double counting we consider only the subprocesses~(1) and~(2).
It is in a contrast with the collinear QCD factorization
where contributions from the subprocesses (1) --- (3) should be taken into account separately.

Let us start from the kinematics.
We denote the 4-momenta of incoming partons and outgoing leptons by $k_1$, $k_2$, $p_1$ and $p_2$. 
The initial hadrons have the 4-momenta $p^{(1)}$ and $p^{(2)}$, and the final 
state quark in~(2) has 4-momentum $p_3$. 
In the center-of-mass frame of colliding particles we can write:
\begin{equation}
p^{(1)} = {\sqrt s\over 2} (1,0,0,1),\quad p^{(2)} = {\sqrt s\over 2} (1,0,0,-1)
\end{equation}

\noindent 
where $\sqrt s$ is the total energy of the process under consideration and we neglect the masses
of the incoming protons. The initial parton four-momenta in the high energy limit can be
written as follows:
\begin{equation}
k_1 = x_1 p^{(1)} + k_{1T},\quad k_2 = x_2 p^{(2)} + k_{2T},
\end{equation}

\noindent 
where $k_{1T}$ and $k_{2T}$ are the corresponding transverse 4-momenta. 
It is important that ${\mathbf k}_{1T}^2 = - k_{1T}^2 \neq 0$, 
${\mathbf k}_{2T}^2 = - k_{2T}^2 \neq 0$. From the conservation laws we can easily obtain 
the following relations for annihilation subprocess~(1):
\begin{equation}
{\mathbf k}_{1T} + {\mathbf k}_{2T} = {\mathbf p}_{1T} + {\mathbf p}_{2T},
\end{equation}
\begin{equation}
x_1 \sqrt s = m_{1T} e^{y_1} + m_{2T} e^{y_2}, 
\end{equation}
\begin{equation}
x_2 \sqrt s = m_{1T} e^{-y_1} + m_{2T} e^{-y_2},
\end{equation}

\noindent 
and the similar ones for QCD Compton subprocess~(2):
\begin{equation}
{\mathbf k}_{1T} + {\mathbf k}_{2T} = {\mathbf p}_{1T} + {\mathbf p}_{2T} + {\mathbf p}_{3T}, 
\end{equation}
\begin{equation}
x_1 \sqrt s = m_{1T} e^{y_1} + m_{2T} e^{y_2} + m_{3T} e^{y_3}, 
\end{equation}
\begin{equation}
x_2 \sqrt s = m_{1T} e^{-y_1} + m_{2T} e^{-y_2} + m_{3T} e^{-y_3},
\end{equation}

\noindent 
where $p_{1T}$, $p_{2T}$ and $p_{3T}$ are the transverse momenta 
of produced particles, $y_1$, $y_2$ and $y_3$ are their center-of-mass rapidities 
and $m_{1T}$, $m_{2T}$ and $m_{3T}$ are the corresponding transverse masses, i.e.
$m_{iT}^2 = m_i^2 + {\mathbf p}_{iT}^2$.
The matrix elements of~(1) and~(2) can be presented as follows:
\begin{equation}
\label{Mqq2G2ll}
\mathcal M_1^\gamma = i e^2 e_q \, \bar v_{s_1}(k_2)\gamma^\mu u_{s_2}(k_1)\,\frac{g_{\mu\nu}}{s}\,\bar u_{r_1}(p_1)\gamma^\nu v_{r_2}(p_2),
\end{equation}
\begin{equation}
\label{Mqq2Z2ll}
\displaystyle \mathcal M_1^Z = i\frac{g^2_w}{4\cos^2\theta_W}\,\bar v_{s_1}(k_2)\gamma^\mu(C_V^q-C_A^q\gamma^5) u_{s_2}(k_1) \times \atop { \displaystyle \times \left( g_{\mu\nu} - {(k_1 + k_2)_\mu (k_1 + k_2)_\nu\over m_Z^2} \right) \, { \bar u_{r_1}(p_1)\gamma^\nu(C_V^e-C_A^e\gamma^5) v_{r_2}(p_2) \over (s-m_Z^2-im_Z\Gamma_Z)},}
\end{equation}
\begin{equation}
\label{Mqg2G2llq}
\displaystyle \mathcal M_2^\gamma=-e^2 e_q g_s t^a \, \epsilon_\mu(k_2) \bar u_{s_1}(k_1)\left(\gamma^\nu\frac{\hat k_1+\hat k_2}{s}\gamma^\mu+\gamma^\mu\frac{-\hat k_2+\hat p_3}{(-k_2+p_3)^2}\gamma^\nu\right)u_{s_2}(p_3) \times \atop { \displaystyle \times \,\frac{g_{\nu\rho}}{(p_1+p_2)^2}\,\bar u_{r_1}(p_1)\gamma^\rho v_{r_2}(p_2) },
\end{equation}
\begin{equation}
\label{Mqg2Z2llq}
\displaystyle \mathcal M_2^Z= -\frac{g_w^2 g_s}{4\cos^2\theta_W}\,t^a \epsilon_\mu(k_2) \times \atop { \displaystyle \times \, \bar u_{s_1}(k_1) \left(\gamma^\nu(C_V^q-C_A^q\gamma^5)\frac{\hat k_1+\hat k_2}{s}\gamma^\mu+\gamma^\mu\frac{-\hat k_2+\hat p_3}{(-k_2+p_3)^2}\gamma^\nu(C_V^q-C_A^q\gamma^5)\right) u_{s_2}(p_3) \times 
  \atop {\displaystyle \times \left( {\displaystyle g_{\rho\nu}-\frac{(p_1+p_2)_\rho(p_1+p_2)_\nu}{m_Z^2}} \right) { \bar u_{r_1}(p_1)\gamma^\rho(C_V^e-C_A^e\gamma^5) v_{r_2}(p_2) \over (p_1+p_2)^2-m_Z^2-im_Z\Gamma_Z}, } }
\end{equation}

\noindent
where $e$ and $e_q$ are the electron and quark (fractional) electric charges, 
$s=(k_1+k_2)^2$, $g_w$ and $g_s$ are the weak and strong charges, $m_Z$ and $\Gamma_Z$ are the mass and full decay width of 
$Z$ boson, $\theta_W$ is the Weinberg mixing angle, $\epsilon^\mu$ 
and $a$ are the polarization 4-vector and eight-fold color index
of incoming off-shell gluon, $C_V$ and $C_A$ are the vector and axial constants.
Here we neglected the masses and virtualities of incoming quarks and took
propagator of intermediate $Z$ boson in a Breit-Wigner form 
to avoid an artificial singularities in numerical calculations.
When we calculate the matrix elements squared, 
the summation over the incoming off-shell gluon polarizations is 
carried out in according to the $k_T$-factorization prescription\cite{24}:
\begin{equation}
\sum \epsilon^\mu \epsilon^{*\, \nu} = {\mathbf k}_{2T}^{\mu} {\mathbf k}_{2T}^{\nu}/{\mathbf k}_{2T}^2.
\end{equation}

\noindent
In the collinear limit, where $|{\mathbf k}_{2T}| \to 0$, this expression converges to the 
ordinary $\sum \epsilon^\mu \epsilon^{*\, \nu} = - g^{\mu \nu}/2$ after averaging on the azimuthal angle.
In all other respects the evaluation follows the standard QCD Feynman rules.
The calculation of traces in (12) --- (15) is straightforward and was done using the algebraic 
manipulation systems {\textsc form}\cite{40}. We do not list here the obvious expressions because of lack of space.
The obtained expression for Compton subprocess~(2) coincides with the result\cite{32}.

To calculate the cross section of Drell-Yan lepton pair production,
in according to the $k_T$-factorization theorem, 
one should convolute off-shell partonic cross sections
with the relevant unintegrated quark and/or gluon distributions in a proton:
\begin{equation}
\label{sigma}
  \sigma = \sum_{i,j = q,\,g} \int {\hat \sigma}_{ij}^*(x_1, x_2, {\mathbf k}_{1T}^2, {\mathbf k}_{2T}^2) \, f_i(x_1,{\mathbf k}_{1T}^2,\mu^2) f_j(x_2,{\mathbf k}_{2T}^2,\mu^2) \, dx_1 dx_2 \, d{\mathbf k}_{1T}^2 d{\mathbf k}_{2T}^2, 
\end{equation}

\noindent
where ${\hat \sigma}_{ij}^*(x_1, x_2, {\mathbf k}_{1T}^2, {\mathbf k}_{2T}^2)$
is the off-shell partonic cross section and $f_i(x,{\mathbf k}_{T}^2,\mu^2)$ is the
unintegrated parton densities in a proton. The contributions to the total Drell-yan cross section from 
quark-antiquark annihilation and QCD Compton subprocesses can be easily rewritten as follows:
\begin{equation}
  \displaystyle \sigma = \sum_{q} \int {1\over 16\pi (x_1 x_2 s)^2 } |\bar {\cal M}_1^{\gamma,\,Z}|^2 \times \atop
  \displaystyle  \times f_q(x_1,{\mathbf k}_{1T}^2,\mu^2) f_q(x_2,{\mathbf k}_{2T}^2,\mu^2) d{\mathbf p}_{1T}^2 d{\mathbf k}_{1T}^2 d{\mathbf k}_{2T}^2 dy_1 dy_2 {d\phi_1 \over 2\pi} {d\phi_2 \over 2\pi},
\end{equation}
\begin{equation}
  \displaystyle \sigma = \sum_{q} \int {1\over 256\pi^3 (x_1 x_2 s)^2 } |\bar {\cal M}_2^{\gamma,\,Z}|^2 \times \atop
  \displaystyle  \times f_q(x_1,{\mathbf k}_{1T}^2,\mu^2) f_g(x_2,{\mathbf k}_{2T}^2,\mu^2) d{\mathbf p}_{1T}^2 d{\mathbf p}_{2T}^2 d{\mathbf k}_{1T}^2 d{\mathbf k}_{2T}^2 dy_1 dy_2 dy_3 {d\phi_1 \over 2\pi} {d\phi_2 \over 2\pi} {d\psi_1 \over 2\pi} {d\psi_2 \over 2\pi},
\end{equation}

\noindent 
where $\phi_1$, $\phi_2$, $\psi_1$ and $\psi_2$ are the azimuthal angles of initial partons
and produced leptons, respectively. If we average these expressions over $\phi_1$ and $\phi_2$ and 
take the limit $|{\mathbf k}_{1T}| \to 0$ and  $|{\mathbf k}_{2T}| \to 0$, then we recover the 
corresponding formulas in the collinear QCD factorization.

Concerning the unintegrated quark and gluon densities in 
a proton, we apply the Kimber-Martin-Ryskin (KMR) approach\cite{41} to calculate them. The KMR approach is the formalism
to construct the unintegrated parton distributions from the known conventional ones. 
In this approximation the unintegrated quark and gluon distributions are given by
\begin{equation}
  \displaystyle f_q(x,{\mathbf k}_T^2,\mu^2) = T_q({\mathbf k}_T^2,\mu^2) {\alpha_s({\mathbf k}_T^2)\over 2\pi} \times \atop {
  \displaystyle \times \int\limits_x^1 dz \left[P_{qq}(z) {x\over z} q\left({x\over z},{\mathbf k}_T^2\right) \Theta\left(\Delta - z\right) + P_{qg}(z) {x\over z} g\left({x\over z},{\mathbf k}_T^2\right) \right],}
\end{equation}
\begin{equation}
  \displaystyle f_g(x,{\mathbf k}_T^2,\mu^2) = T_g({\mathbf k}_T^2,\mu^2) {\alpha_s({\mathbf k}_T^2)\over 2\pi} \times \atop {
  \displaystyle \times \int\limits_x^1 dz \left[\sum_q P_{gq}(z) {x\over z} q\left({x\over z},{\mathbf k}_T^2\right) + P_{gg}(z) {x\over z} g\left({x\over z},{\mathbf k}_T^2\right)\Theta\left(\Delta - z\right) \right],} 
\end{equation}

\noindent
where $P_{ab}(z)$ are the usual unregulated LO DGLAP splitting 
functions. The theta functions which appears
in~(20) and~(21) imply the angular-ordering constraint $\Delta = \mu/(\mu + |{\mathbf k}_T|)$ 
specifically to the last evolution step to regulate the soft gluon
singularities. 
Numerically, for the input we have used leading-order parton densities $xq(x,\mu^2)$ and 
$xg(x,\mu^2)$ from recent MSTW'2008 set\cite{42}. 
The Sudakov form factors $T_q({\mathbf k}_T^2,\mu^2)$ and 
$T_g({\mathbf k}_T^2,\mu^2)$ enable us to include logarithmic loop corrections
to the calculated cross sections. To take into account the non-logarithmic loop corrections 
we use the approach proposed in\cite{43}.
It was demonstrated that main part of the non-logarithmic loop corrections to the 
quark-antiquark annihilation cross section~(1) can be 
absorbed in the effective $K$-factor:
\begin{equation}
  K = \exp \left[ C_F {\alpha_s(\mu^2)\over 2\pi } \pi^2 \right],
\end{equation}

\noindent 
where color factor $C_F = 4/3$. A particular choice $\mu^2 = {\mathbf p}_T^{4/3} M^{2/3}$ has been 
proposed\cite{23,43} to eliminate sub-leading logarithmic terms. We choose this scale to evaluate the strong
coupling constant in~(22).

The multidimensional integrations in~(18) and~(19) have been performed
by the means of Monte Carlo technique, using the routine \textsc{vegas}\cite{44}.
The full C$++$ code is available from the author on 
request\footnote{lipatov@theory.sinp.msu.ru}.

\section{Numerical results} \indent

We now are in a position to present our numerical results. First we describe our
input and the kinematic conditions. After we fixed the unintegrated
gluon distributions, the cross sections (18) and (19) depend on
the renormalization and factorization scales $\mu_R$ and $\mu_F$. 
Numerically, we set them to be equal to $\mu_R = \mu_F = \xi M$. To estimate the scale 
uncertainties of our calculations
we vary the parameter $\xi$ between 1/2 and 2 about the default value $\xi = 1$.
Following to\cite{45}, we set $m_Z = 91.1876$~GeV, $\Gamma_Z = 2.4952$~GeV,
$\sin^2 \theta_W=0.23122$
and use the LO formula for the strong 
coupling constant $\alpha_s(\mu^2)$ with $n_f = 4$ 
active quark flavors at $\Lambda_{\rm QCD} = 200$~MeV, so that $\alpha_s(M_Z^2) = 0.1232$.

The results of our calculations are presented in Figs.~2 --- 4 in 
comparison with the D$\emptyset$\cite{38}, CDF\cite{33,34,35,36,37} and CMS data\cite{39}. 
Solid histograms are obtained by fixing both the
factorization and renormalization scales at the default value $\mu = M$,
whereas the upper and lower dashed histograms correspond to the scale variation as it
was described above. The predicted total cross sections are listed in Table~1.
One can see that the Tevatron and LHC experimental data are reasonable well described by
the $k_T$-factorization approach in the whole range of invariant masses. 
Our predictions tend to only slightly overestimate the rapidity distribution
of dilepton pair in the region of $Z$ boson peak $66 < M < 116$~GeV, but agree
with the data within the uncertainties. Specially we point out a good description of dilepton 
transverse momentum distributions measured by the CDF collaboration since this observable strongly
depends on the unintegrated parton density used. 

The relative contributions of quark-antiquark annihilation and QCD Compton subprocesses
to the Drell-Yan cross sections at the Tevatron and LHC energies are shown in Fig.~5 as a function of azimuthal angle difference 
between the transverse momenta of produced leptons.
Note that this observable is singular in the collinear QCD approximation at LO due to back-to-back
kinematics. It is in a contrast with the $k_T$-factorization approach, where, as it was mentioned above, 
the finite transverse momentum of dilepton pair 
is generated already in LO quark-antiquark annihilation~(1).
We find that latter dominates at high $\Delta \phi \sim \pi$ for both the Tevatron and LHC energies,
whereas at $\Delta \phi < \pi/2$ quark-antiquark annihilation and QCD Compton subprocesses contribute equally.
Note that here we applied no cuts on the final-state phase space.

\begin{table}
\begin{center}
\begin{tabular}{|l|c|c|}
\hline
   & &\\
  Source & $\sigma(66 < M < 116$~GeV), pb & $\sigma(M > 116$~GeV), pb \\
   & &\\
\hline
   & &\\
   $k_T$-factorization (KMR) & $285 \pm 31$ & $3.7 \pm 0.4$\\
   & &\\
   NNLO pQCD\cite{36} & $227\pm9$ & 3.3\\
   & &\\
   CDF data \cite{36} & $250 \pm 4$ (stat.) $ \pm 10$ (syst.) & $4.0 \pm 0.4$ (stat.+syst.) $ \pm 0.2$ (lumi.) \\
   & &\\
\hline
\end{tabular}
\end{center}
\caption{Total cross sections of Drell-Yan pair production in $p\bar p$ collisions at $\sqrt s = 1800$~GeV.}
\end{table}

Now we turn to more detailed analysis of angular distributions in dilepton production.
The general expression can be described by the polar $\theta$ and azimuthal 
$\phi$ angles of produced particles in the dilepton rest frame. 
When integrated over $\cos \theta$ or $\phi$, respectively, the angular distribution can be presented as follows:
\begin{equation}
  {d\sigma \over d\cos \theta} \sim (1 + \cos^2\theta) + {1\over 2} A_0 (1 - 3\cos^2\theta) + A_4\cos\theta,
\end{equation}
\begin{equation}
  {d\sigma \over d \phi} \sim 1 + \beta_3\cos\phi + \beta_2\cos2\phi,
\end{equation}

\noindent
where $\beta_3 = 3\pi A_3/16$ and $\beta_2 = A_2/4$. 
Note that the angular coefficients $A_0$ and $A_2$ are the same for $\gamma^*$ or $Z$ boson exchange, and $A_3$ and $A_4$ 
originate from the $\gamma^* - Z$ interference.
The Lam-Tung relation\cite{46} $A_0 = A_2$ is valid for both quark-antiquark annihilation and QCD Compton subprocesses
at ${\cal O}(\alpha \alpha_s)$ order. Higher-order QCD calculations\cite{47,48} as 
well as QCD resummation up to all orders\cite{49}
indicate that violations of the Lam-Tung relation are small. Very recently
the CDF collaboration reported\cite{37} the first measurement of the angular 
coefficients $A_0, A_2, A_3$ and $A_4$ in the $Z$ peak region ($66 < M < 116$~GeV) at $\sqrt s = 1960$~GeV.
Below we estimate these coefficients regarding the CDF measurements.
Our evaluation generally followed the experimental
procedure. We have collected the simulated
events in the specified bins of dilepton transverse momentum,
generated the decay lepton angular distributions according
to the matrix elements (12) --- (15) and then applied a two-parametric fit based on~(23) and~(24).
The estimated values of angular coefficients in the Collins-Soper frame are shown in Fig.~6.
We find that our predictions agree well with the CDF data as well as collinear QCD predictions listed in\cite{37}. 
We would like to only remark that the latter
predict a flat behaviour of $A_3$ in a whole $p_T$ range whereas CDF data tends to support 
our predictions (slight decreasing of $A_3$ when we move to large $p_T$ values).

Finally, we can conclude that $k_T$-factorization predictions in general are rather similar to ones based 
on the collinear QCD factorization with the NNLO accuracy.
It demonstrates again that the $k_T$-factorization approach at LO level automatically 
incorporates a large piece of the standard (collinear) high-order corrections\cite{27}.
It is important for further studies of small-$x$ physics at hadron colliders, 
and, in particular, for searches of effects of new physics beyond the SM at the LHC.

\section{Conclusions}

We have investigated unpolarized Drell-Yan lepton pair production in $p\bar p$ and $pp$ 
collisions at the Tevatron and LHC energies within the framework of the $k_T$-factorization approach.
Our consideration is based on the ${\cal O}(\alpha)$ and ${\cal O}(\alpha \alpha_s)$ off-shell 
production amplitudes where $\gamma^*$ and $Z$ boson exchange is taken into account.
The calculations include leptonic decays of $Z$ bosons 
with full spin correlations and $\gamma^* - Z$ interference. 
The unintegrated parton densities in a proton are determined by the 
Kimber-Martin-Ryskin prescription. 
We obtained a reasonable well agreement 
(at a similar level as in the 
NNLO pQCD) between our predictions and the available data 
taken by the D$\emptyset$, CDF and CMS collaborations.
Specially we studied the specific angular distributions measured very recently by the CDF collaboration
for the first time.

\section{Acknowledgments} \indent 

We thank S.P.~Baranov, L.N.~Lipatov and A.~Szczurek
for encouraging interest and useful discussions.
A.V.L. and N.P.Z. are very grateful to 
DESY Directorate for the support in the 
framework of Moscow --- DESY project on Monte-Carlo
implementation for HERA --- LHC.
A.V.L. and M.A.M. were supported in part by the grant of president of 
Russian Federation (MK-3977.2011.2).
Also this research was supported by the 
FASI of Russian Federation (grant NS-4142.2010.2),
FASI state contract 02.740.11.0244, 
RFBR grant 11-02-01454-a and the RMES (grant the Scientific Research on High Energy Physics).

\newpage

\begin{figure}
\begin{center}
\epsfig{figure=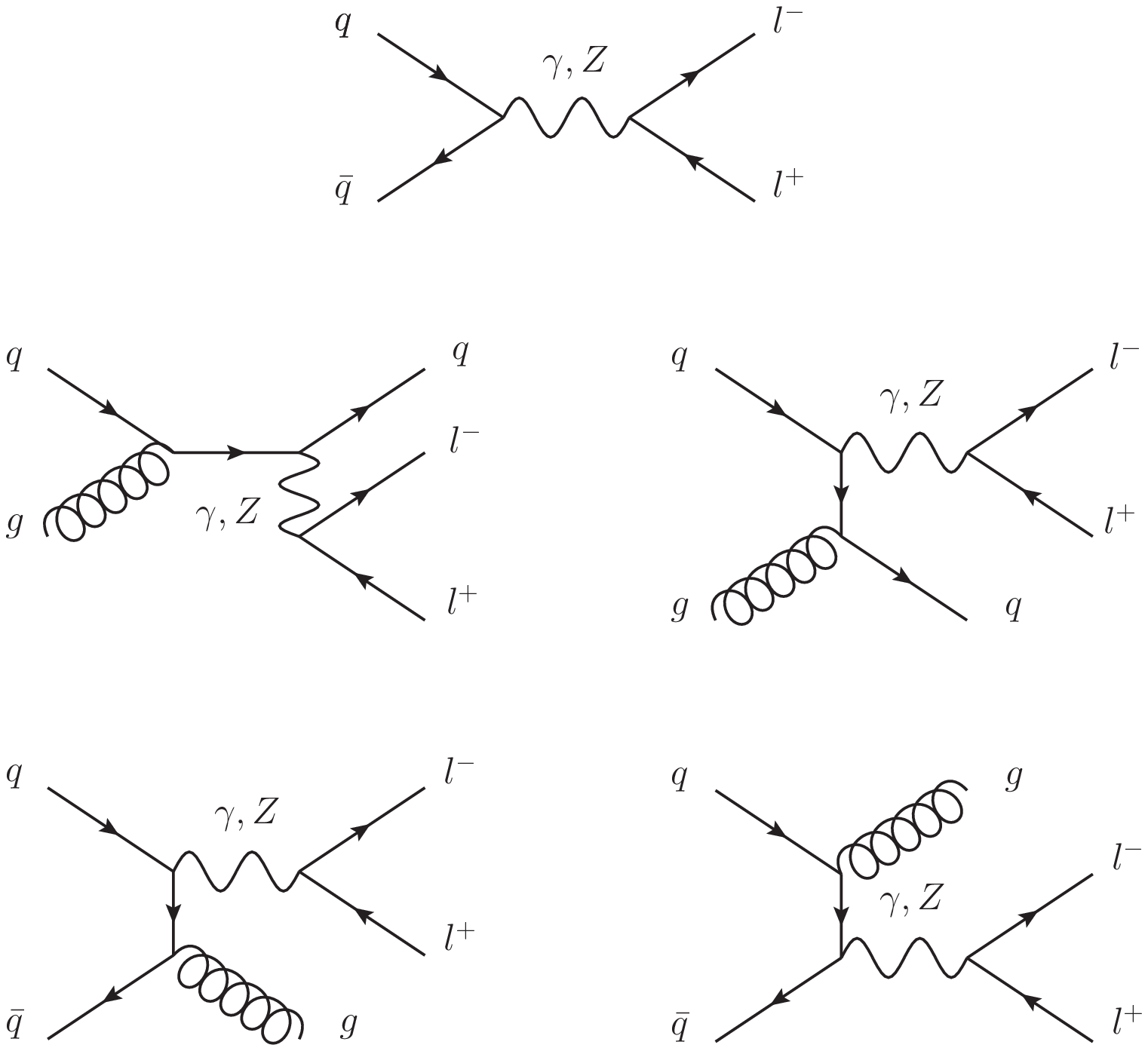, width = 12cm}
\vspace{1.5cm}
\caption{The Feynman graphs for Drell-Yan pair production at the ${\cal O}(\alpha)$ and ${\cal O}(\alpha \alpha_s)$ orders.}
\label{fig1}
\end{center}
\end{figure}

\begin{figure}
\begin{center}
\epsfig{figure=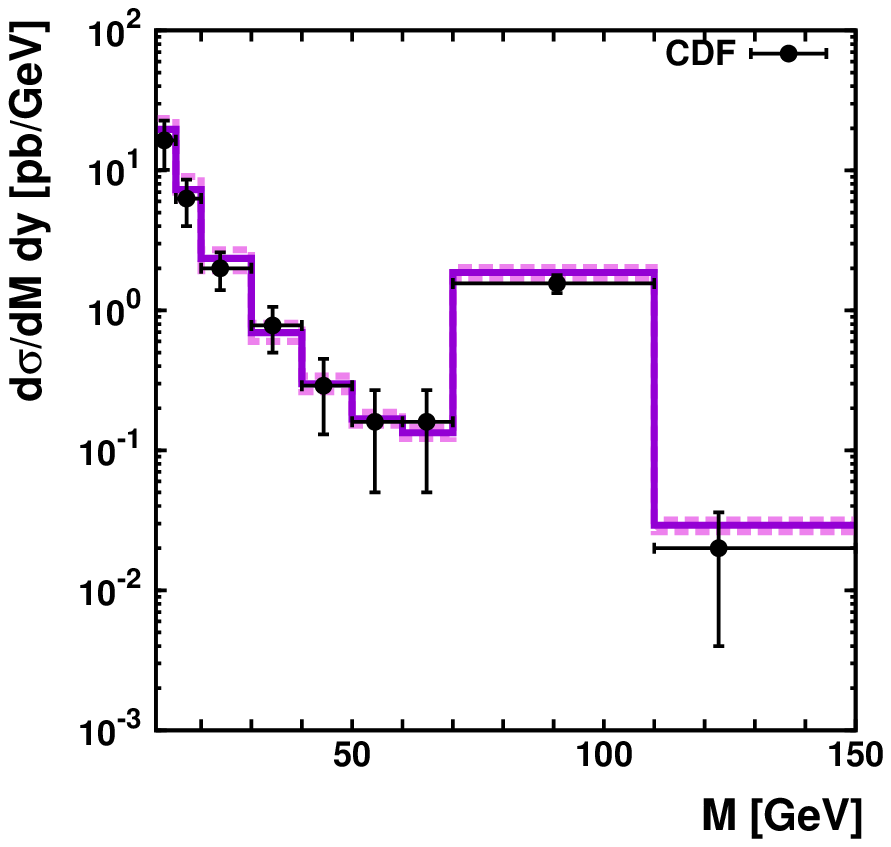, width = 8.1cm}
\epsfig{figure=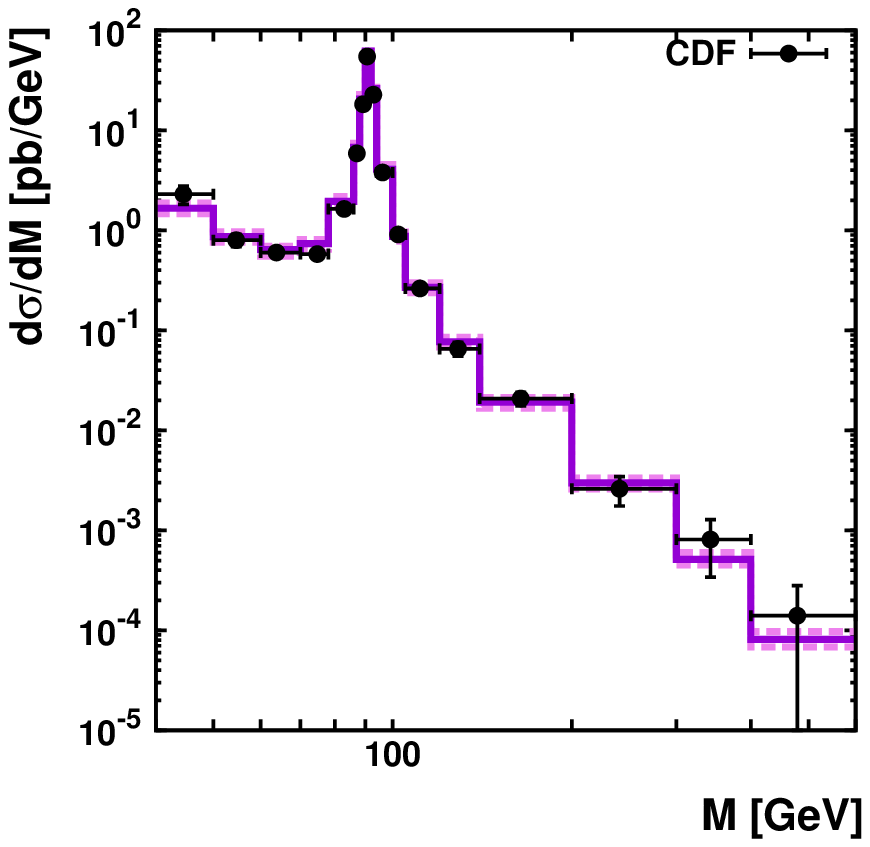, width = 8.1cm}
\epsfig{figure=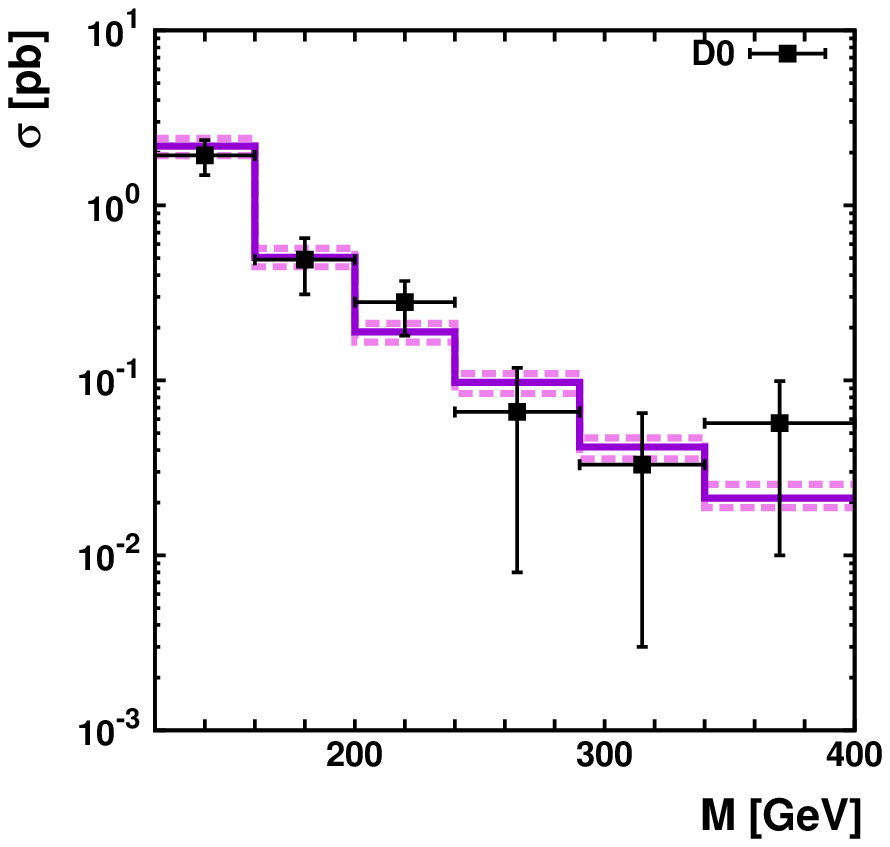, width = 8.1cm}
\epsfig{figure=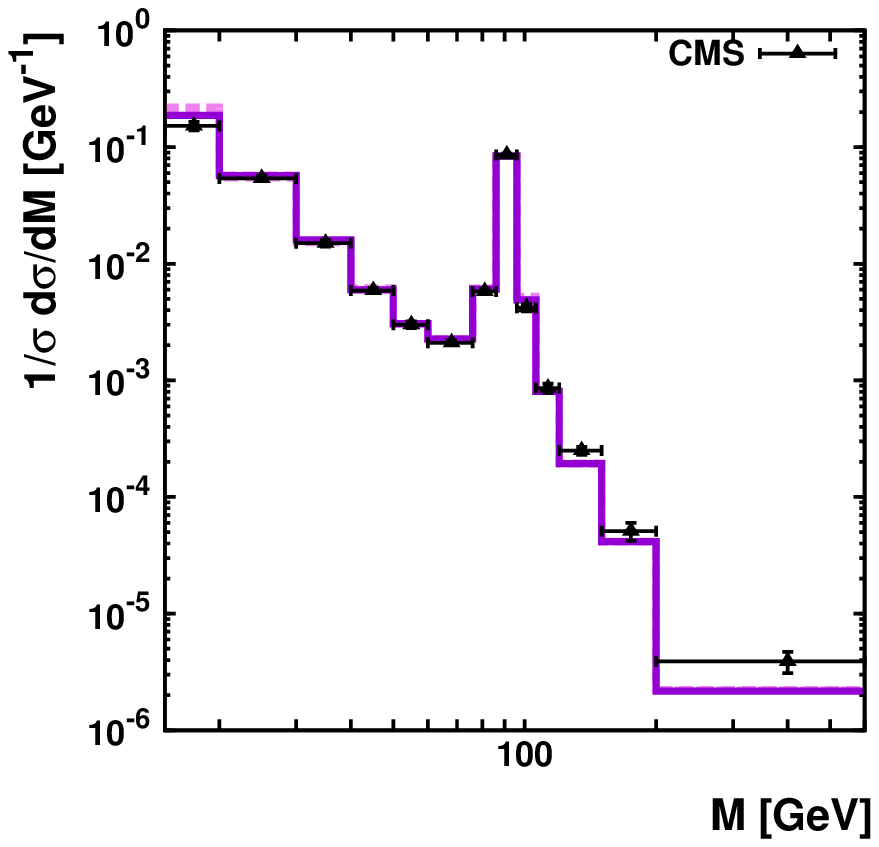, width = 8.1cm}
\caption{The total and differential cross sections of the 
Drell-Yan pair production in $p\bar p$ and $pp$ collisions
at the Tevatron and LHC as a function of dilepton invariant mass $M$.
The solid histograms correspond to the results obtained with the KMR parton densities.
The upper and lower dashed histograms correspond to scale variations, as it is
described in the text. The experimental data are from D$\emptyset$\cite{38}, CDF\cite{33,35} and CMS\cite{39}.}
\label{fig2}
\end{center}
\end{figure}

\begin{figure}
\begin{center}
\epsfig{figure=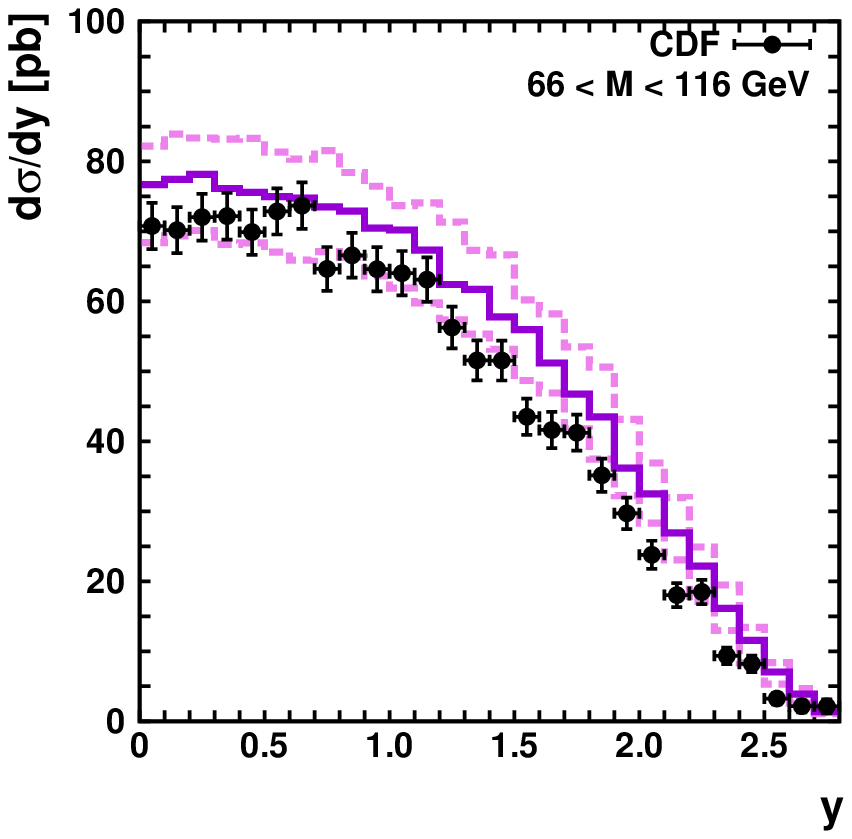, width = 8.1cm}
\epsfig{figure=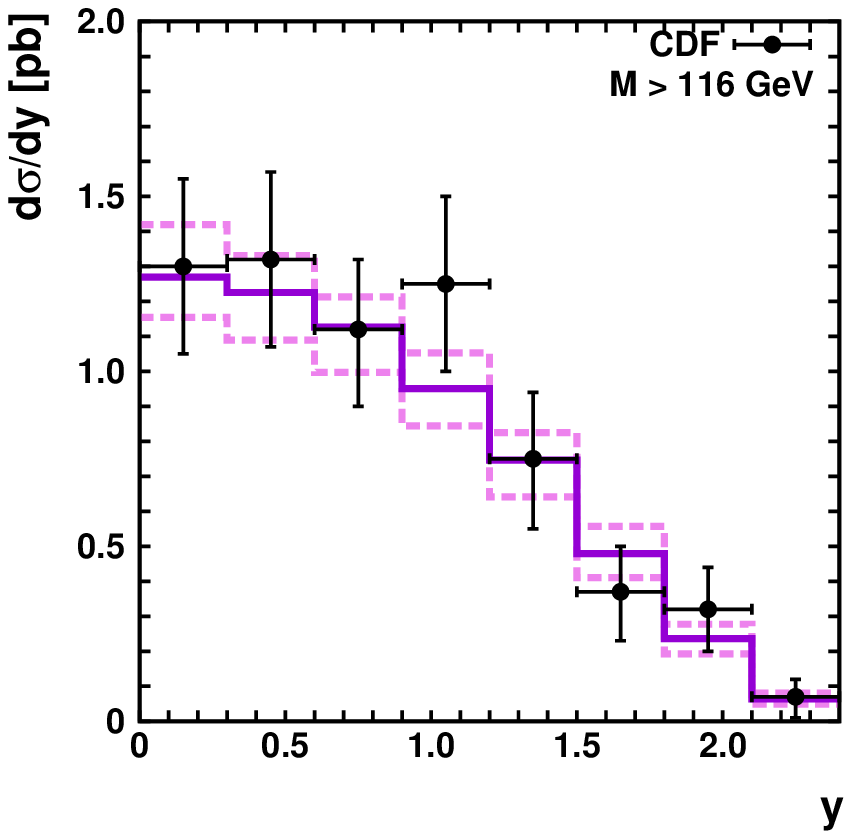, width = 8.1cm}
\caption{The differential cross sections $d\sigma/dy$ of dilepton production
at $\sqrt s = 1800$~TeV compared to the CDF data\cite{36}. 
Notation of all histograms is the same as in Fig.~1.}
\label{fig3}
\end{center}
\end{figure}

\begin{figure}
\begin{center}
\epsfig{figure=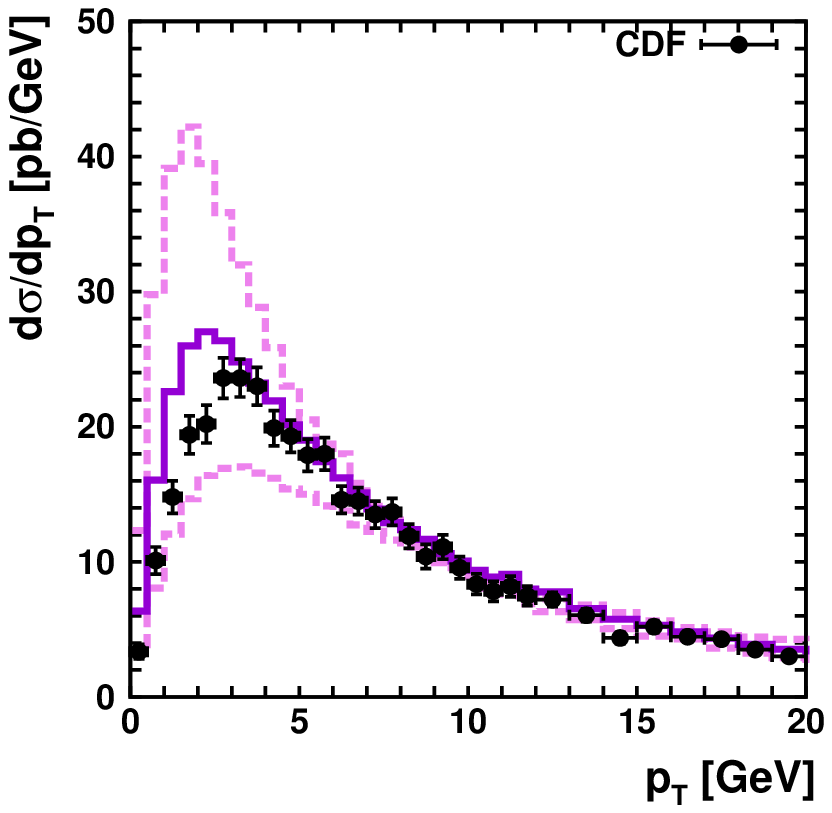, width = 8.1cm}
\epsfig{figure=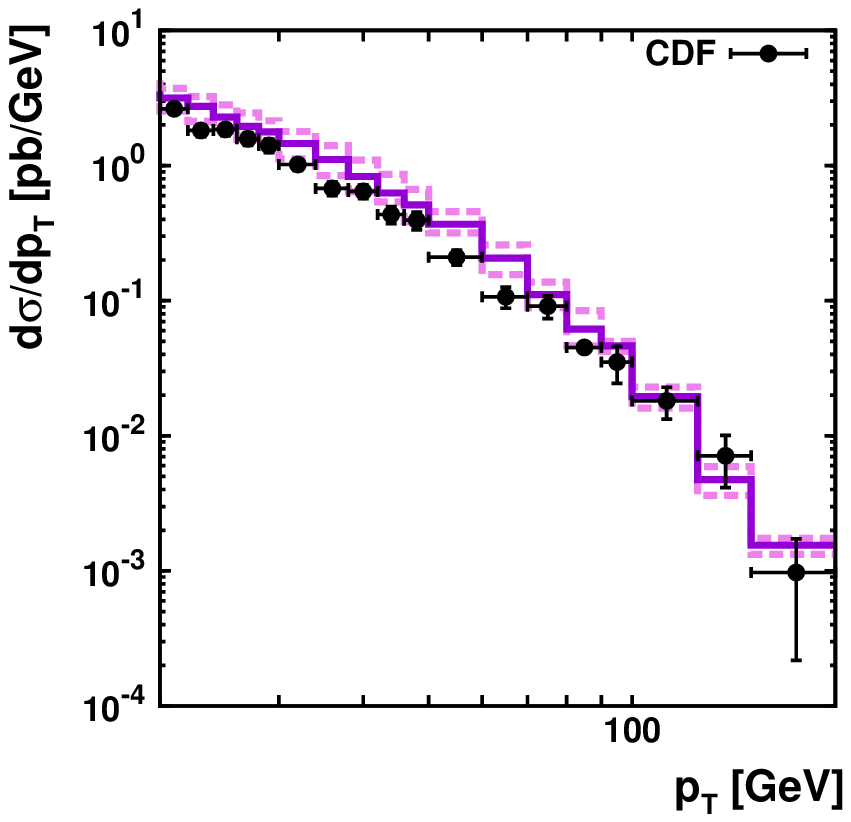, width = 8.1cm}
\caption{The differential cross sections $d\sigma/dp_T$ of dilepton production
at $\sqrt s = 1800$~TeV compared to the CDF data\cite{34}. 
Notation of all histograms is the same as in Fig.~1.}
\label{fig4}
\end{center}
\end{figure}

\begin{figure}
\begin{center}
\epsfig{figure=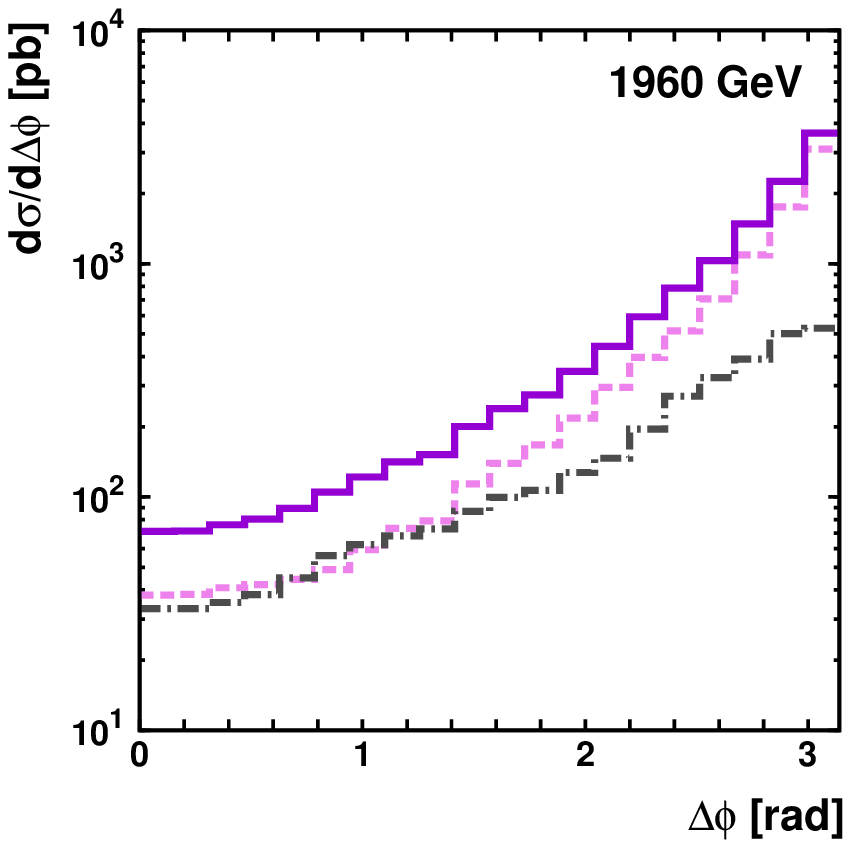, width = 8.1cm}
\epsfig{figure=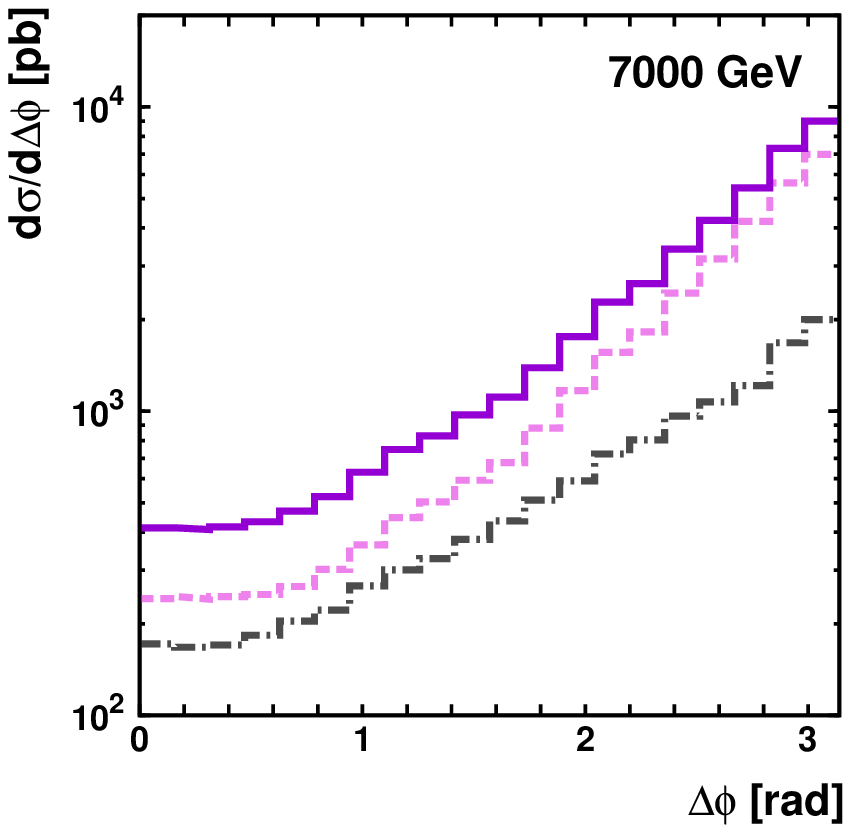, width = 8.1cm}
\caption{Different contributions to the Drell-Yan pair cross section in $p \bar p$ and $pp$ 
collisions at the Tevatron and LHC energies as a function of produced lepton azimuthal angle difference.
The dashed and dash-dotted histograms correspond to the
contributions from quark-antiquark annihilation and QCD Compton subprocesses, respectively. 
The solid histograms represent the sum of these components.}
\label{fig5}
\end{center}
\end{figure}

\begin{figure}
\begin{center}
\epsfig{figure=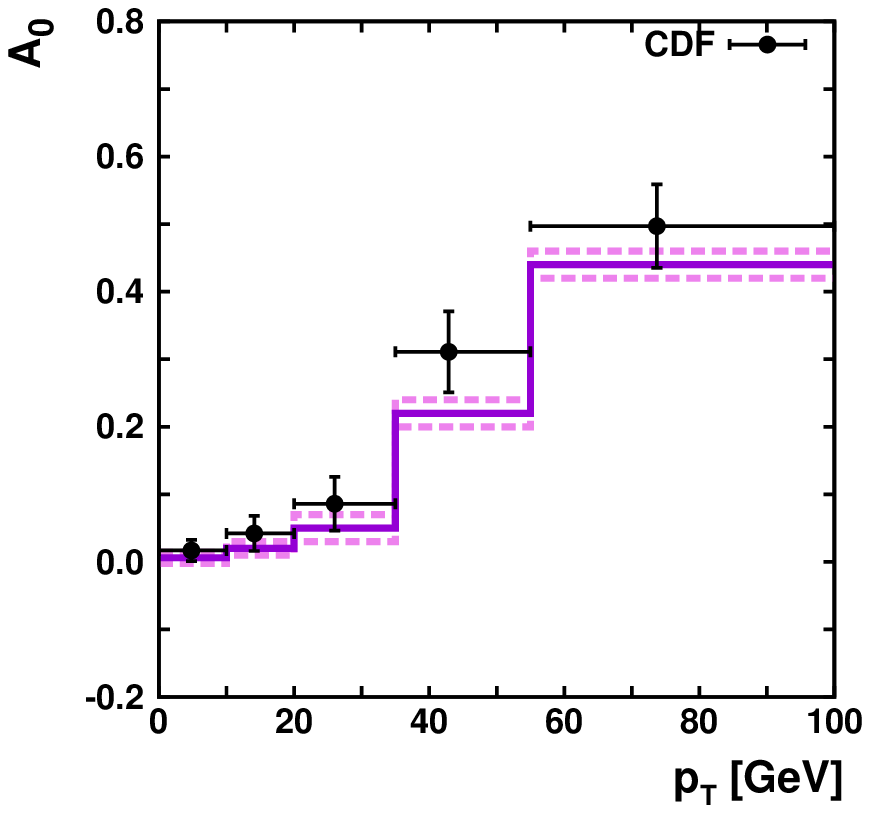, width = 8.1cm}
\epsfig{figure=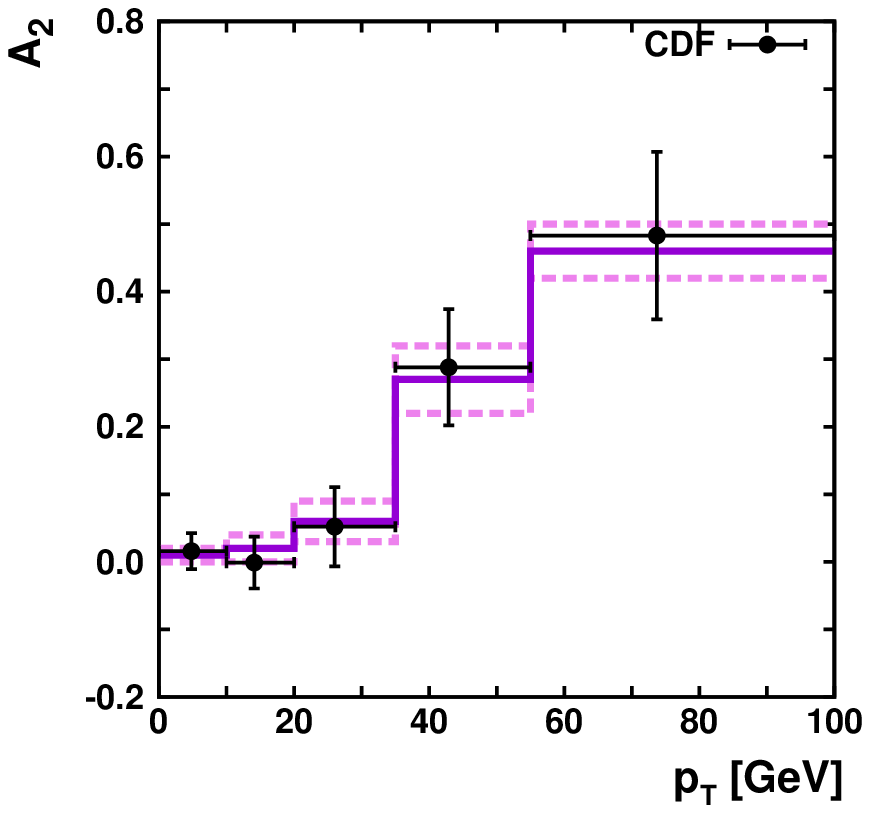, width = 8.1cm}
\epsfig{figure=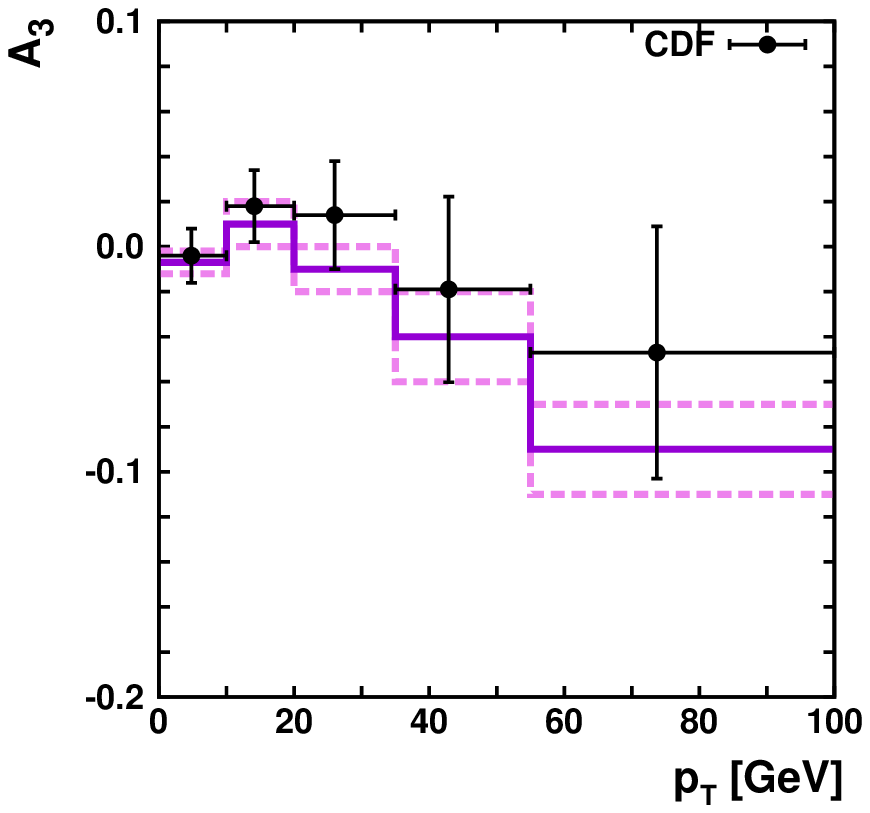, width = 8.1cm}
\epsfig{figure=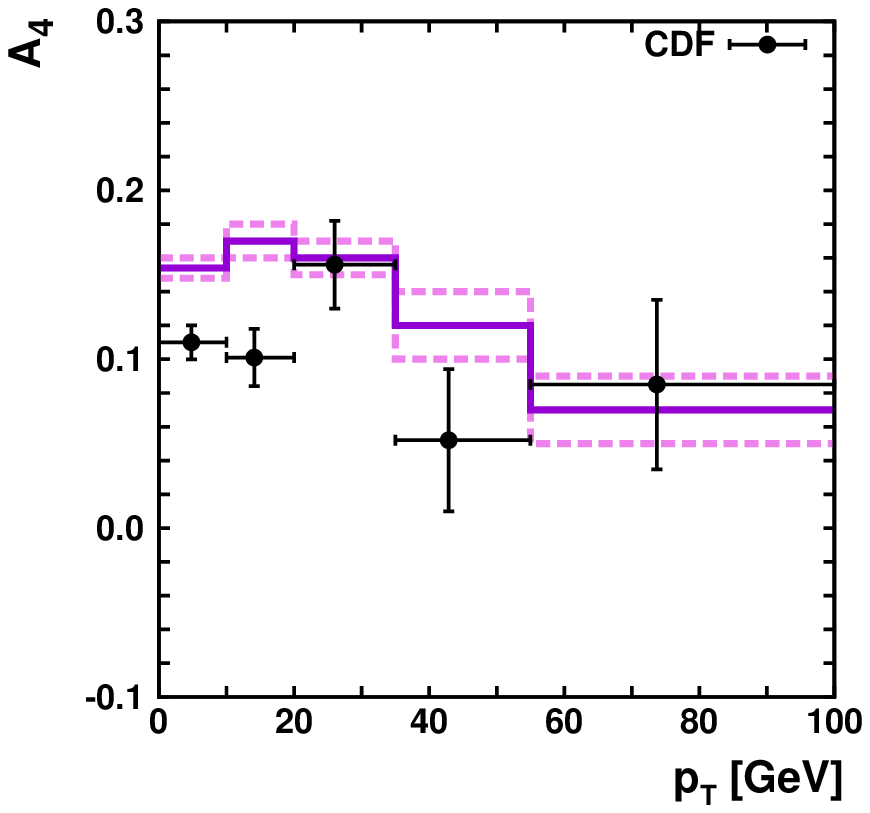, width = 8.1cm}
\caption{Angular coefficients $A_0$, $A_2$, $A_3$ and $A_4$ of dilepton production
as a function of $p_T$ compared to the CDF data\cite{37}. 
Solid and two dashed histograms represent fitted values of
angular coefficients and corresponding uncertainties of fitting procedure.
The default scale $\mu = M$ has been applied.}
\label{fig6}
\end{center}
\end{figure}

\end{document}